**Title: Should we build more large dams? The actual costs of hydropower megaproject development**

**Authors:** Atif Ansar[1,2,*], Bent Flyvbjerg[1], Alexander Budzier[1], Daniel Lunn[3]

**Affiliations:**
[1]Saïd Business School, University of Oxford, OX1 1HP, UK.
[2]Blavatnik School of Government, University of Oxford, OX1 4JJ, UK.
[3]Department of Statistics, University of Oxford, OX1 3GT, UK.

* To whom correspondence should be addressed.
E-mail: atif.ansar@bsg.ox.ac.uk
Tel: +44 (0) 1865 616690

**Abstract**: A brisk building boom of hydropower mega-dams is underway from China to Brazil. Whether benefits of new dams will outweigh costs remains unresolved despite contentious debates. We investigate this question with the "outside view" or "reference class forecasting" based on literature on decision-making under uncertainty in psychology. We find overwhelming evidence that budgets are systematically biased below actual costs of large hydropower dams—excluding inflation, substantial debt servicing, environmental, and social costs. Using the largest and most reliable reference data of its kind and multilevel statistical techniques applied to large dams for the first time, we were successful in fitting parsimonious models to predict cost and schedule overruns. The outside view suggests that in most countries large hydropower dams will be too costly in absolute terms and take too long to build to deliver a positive risk-adjusted return unless suitable risk management measures outlined in this paper can be affordably provided. Policymakers, particularly in developing countries, are advised to prefer agile energy alternatives that can be built over shorter time horizons to energy megaprojects.

**Main Text:**

**1. Large hydropower dam controversy**

The 21st Century faces significant energy challenges on a global scale. Population and economic growth underpin increasing demand for energy from electricity to transport fuels. Social objectives of poverty alleviation, adaptation and mitigation of climate change, and energy security present policy makers and business leaders with difficult decisions and critical trade-offs in implementing sound energy policies. Demand for electricity is, for example, slated to almost double between 2010 and 2035 requiring global electricity capacity to increase from 5.2 terawatt (TW) to 9.3 TW over the same period (IEA, 2011). Currently, the de facto strategic response to these big energy challenges is "big solutions" such as large hydropower dams. Are such big solutions in general and large hydropower dams in particular the most effective strategy, on a risk-adjusted basis, to resolve global energy challenges? Might more numerous small interventions be more prudent from the perspective of risk management and maximizing net present value even when they entail somewhat higher per unit cost of production?

Proponents of large dams envisage multiple benefits. A big step-up in hydropower capacity along with a long and varied list of corollary benefits: reducing fossil fuel consumption, flood control, irrigation, urban water supply, inland water transport, technological progress, and job creation (Billington and Jackson, 2006; ICOLD, 2010). Inspired by the promise of prosperity, there is a robust pipeline of new mega-dams being developed globally after a two-decade lull. The Belo Monte dam in Brazil, the Diamer-Bhasha in Pakistan, Jinsha river dams in China, Myitsone dam in Myanmar, or the Gilgel Gibe III dam in Ethiopia, all in various stages of development, are unprecedented in scale.

Large dams are, however, controversial because they exert substantial financial costs (World Bank, 1996; World Commission on Dams, 2000). Beyond the financial calculus, large dams have profound environmental (McCully, 2001; Scudder, 2005; Stone, 2011), ecological (Nilsson et al., 2005; Ziv et al., 2012), and social (Bakker, 1999; Duflo and Pande, 2007; Richter et al., 2010; Sovacool and Bulan, 2011) impacts. Stone (2011, p. 817) reports in *Science* that the Three Gorges dam in China is an "environmental bane" that will cost over USD 26.45 billion over the next 10-years in environmental "mitigation efforts". Despite their outsized financial and environmental costs, the purported benefits of large hydropower prove uncertain. For example, the World Commission of Dams (2000, p. 30) reported that for large hydropower dams "average [hydropower] generation in the first year of commercial operation is 80% of the targeted value"—a trend of which the recently completed Bakun hydroelectric project in Borneo is an alarming example (Sovacool and Bulan, 2011). Similarly, Duflo and Pande (2007) find adverse distributional impacts of large irrigation dams in India. Winners downstream come with losers upstream yielding a more modest, if any, net economic benefit.

The scale of contemporary dams is so vast that even for a large economy such as China's the negative economic ramifications "could likely hinder the economic viability of the country as a whole" if the risks inherent to these projects are not well managed (Salazar, 2000). Similarly, Merrow et al. (1988, pp. 2-3) warn that "such enormous sums of money ride on the success of megaprojects [such as large dams] that company balance sheets and even government balance-of-payments accounts can be affected for years by the outcomes". Such warnings are not idle alarmism. There is mounting evidence in civil society, academic research, and institutional accounts that large dams have strikingly poor performance records in terms of economy, social and environmental impact, and public support (McCully, 2001; Scudder, 2005; Singh, 2002; Sovacool and Bulan, 2011; WCD, 2000). There are acrimonious, and as yet inconclusive, debates in scientific literature and civil society about whether large dams are a boon or a curse. Should we build more large hydropower dams? How confident can planners be that a large bet on a large dam will pay-off handsomely?

We investigate these questions with the "outside view" or "reference class forecasting" based on the literature on decision-making under uncertainty that won Princeton psychologist Daniel Kahneman the Nobel Prize in economics in 2002 (Kahneman and Tversky, 1979a, 1979b; Kahneman, 1994) extended and applied by Bent Flyvbjerg and colleagues to infrastructure projects (Flyvbjerg et al., 2003; Flyvbjerg, 2009). We present statistical and comparative evidence from the largest reference class of actual costs of large hydropower dam projects (hereafter large dams unless stated otherwise). We find that even before accounting for negative impacts on human society and environment, the actual construction costs of large dams are too high to yield a positive return. Large dams also take inordinately long periods of time to build, making them ineffective in resolving urgent energy crises. Our evidence pertains primarily to large dams and the results cannot be applied either to smaller dams or other large energy solutions such as nuclear power without first conducting a separate "reference class" for other type of power generation technologies. Our findings, however, point towards the generalizable policy proposition that policymakers should prefer energy alternatives that require less upfront outlays and that can be built very quickly.

There is no doubt that harnessing and managing the power of water is critical for economies but large dams are not the way to do so unless suitable risk management measures outlined in this paper can be affordably provided. Building on literature in decision making under uncertainty in management, psychology, and planning research, this paper further provides public agencies (e.g. national planning and finance ministries, power and water authorities), private entrepreneurs, investors, and civil society a framework to test the reliability of *ex ante* estimates for construction costs and schedules of power generation alternatives. An impartial and rigorous application of the reference class forecasting methods proposed here can improve the selection and implementation of new investments.

**2. Delusion and deception in large hydropower dam planning?**

Our approach to address the debates about whether or not to build dams is to incorporate an evidence-based perspective that reflects how decisions among alternative options are actually made and on what basis. Theoretical and empirical literature on decision-making under uncertainty proposes two explanations—psychological delusion and political deception—that suggest decision-makers' forecasts, and hence ex ante judgments, are often adversely biased (Tversky and Kahneman, 1974; Kahneman and Lovallo, 1993; Flyvbjerg, 2003; Lovallo and Kahneman, 2003; Kahneman, 2011).

First, experts (e.g., statisticians, engineers, or economists) and laypersons are systematically and predictably too optimistic about the time, costs, and benefits of a decision. This "planning fallacy" (Kahneman and Tversky, 1979b; Buehler et al., 1994) stems from actors taking an "inside view" focusing on the constituents of the specific planned action rather than on the outcomes of similar actions already completed (Kahneman and Lovallo, 1993). Thus, for example, the estimated costs put forward by cities competing to hold the Olympic Games have consistently been underestimated yet every four years these errors are repeated. Biases, such as overconfidence or overreliance on heuristics (rules-of-thumb) underpin these errors.

Second, optimistic judgments are often exacerbated by deception, i.e. strategic misrepresentation by project promoters (Wachs, 1989; Pickrell, 1992; Flyvbjerg et al., 2002, 2005, 2009). Recent literature on infrastructure delivery finds strong evidence that misplaced political incentives and agency problems lead to flawed decision-making (see Flyvbjerg et al., 2009). Flyvbjerg et al. (2009, p. 180) further discuss that delusion and deception are complementary rather than alternative explanations for why megaprojects typically face adverse outcomes. It is, however, "difficult to disentangle" delusion from deception in practice. Using quasi-experimental evidence from China, Ansar et al. (2013) suggest that while better incentive alignment can help lower the frequency and to a lesser extent the magnitude of biases, it does not entirely cure biases.

Be it delusion or deception, is decision-making in large hydropower dams systematically biased by errors in cost, schedule, and benefit forecasts? What is the risk that costs might outweigh benefits for a proposed dam? While the future is unknowable, uncertain outcomes of large investments can still be empirically investigated using "reference class forecasting" (RCF) or the "outside view" techniques (Kahneman and Lovallo, 1993; Flyvbjerg, 2006, 2008) . To take an outside view on the outcome of an action (or event) is to place it in the statistical distribution of the outcomes of comparable, already-concluded, actions (or events). The outside view has three advantages: First, it is evidence-based and requires no restrictive assumptions. Second, it helps to test and fit models to explain why the outcomes of a reference class of past actions follow the observed distribution. Third, it allows to predict the uncertain outcomes of a planned action by comparing it with the distributional information of the relevant reference class. The theoretical foundations of the outside view were first described by Kahneman and Tversky (1979b) and later by Kahneman and Lovallo (1993) and Lovallo and Kahneman (2003) as a means to detect and cure biases in human judgment. The methodology and data needed for employing the outside view, or reference class forecasting, in practice were developed by Flyvbjerg (2006, 2008) in collaboration with first implemented in practice by Flyvbjerg and COWI, 2004).

*2.1 Three steps to the outside view*

The outside view, applied to large dams for the first time here, involves three steps: i) identify a reference class; ii) establish an empirical distribution for the selected reference class of the parameter that is being forecasted; iii) compare the specific case with the reference class distribution. We take a further innovatory step of fitting multivariate multilevel models to the reference data to predict future outcomes. Our technique is an important improvement in the methodology of the outside view that can be generalized and applied to other large-scale and long-term decisions under uncertainty. With debiased forecasts managers can make statistically grounded, rather than optimistic, judgments (Dawes et al., 1989; Buehler et al., 1994; Gilovich et al., 2002).

The outside view—as implemented by Flyvbjerg (2006, 2008)—is not without its limitations (see Sovacool and Cooper, 2013 for a discussion specifically about energy megaprojects). For example, RCF focuses on generic risk inherent in a reference class rather than specific project-level risk. We rectify against this limitation by fitting regression models in addition to using traditional RCF methods in the result section below. Sovacool and Cooper (2013, p. 63) further suggest that RCF may not provide sufficiently accurate indication of the risks of rare megaprojects the likes of which have never been built before. Such "out of the sample" problems are well noted in probability theory. They do not, however, deny the fundamental usefulness of RCF. If any-thing our results err towards conservative estimates of actual cost overruns and risks experienced by large dams.

*2.2 Measures and data*

Following literature on the planning fallacy (op. cit.), the parameters central to our investigation and multilevel regression analysis is the inaccuracy between managers' forecasts and actual outcomes related to construction costs, or the cost overrun, and implementation schedule, or schedule slippage. Following convention, cost overrun is the actual outturn costs expressed as a ratio of estimated costs[1]; cost overruns can also be thought as the underestimation of actual costs (Bacon and Besant-Jones, 1998; Flyvbjerg et al., 2002). Schedule slippage, called schedule overrun, is the ratio of the actual project implementation duration to the estimated project implementation. The start of the implementation period is taken to be the date of project approval by the main financiers and the key decision makers, and the end is the date of full commercial operation.

Inaccuracies between actual outcomes versus planned forecasts are useful proxies for the underlying risk factors that led to the inaccuracies. For example, cost overruns reduce the attractiveness of an investment and if they become large the fundamental economic viability becomes questionable. Bacon and Besant-Jones (1998, p. 317) offer an astute summary:

> The economic impact of a construction cost overrun is the possible
> loss of the economic justification for the project. A cost overrun
> can also be critical to policies for pricing electricity on the basis of
> economic costs, because such overruns would lead to underpric-

> ing. The financial impact of a cost overrun is the strain on the power utility and on national financing capacity in terms of foreign borrowings and domestic credit.

Similarly, schedule slippages delay much needed benefits, expose projects to risks such as an increase in finance charges, or creeping inflation, which may all require upward revision in the nominal electricity tariffs. Financial costs and implementation schedules, because of their tangibility, are also good proxies for non-pecuniary impacts such as those on the environment or society. Projects with a poor cost and schedule performance are also likely to have a poor environmental and social track record. A greater magnitude of cost and schedule overruns is thus a robust indicator of project failure (Flyvbjerg et a., 2003).

In taking the outside view on the cost and schedule under/overruns, our first step was to establish a valid and reliable reference class of previously built hydropower dams as discussed above. The suggested practice is that a reference class ought to be broad and large enough to be statistically meaningful but narrow enough to be comparable (Kahneman and Tversky, 1979b; Kahneman and Lovallo, 1993; Flyvbjerg, 2006). International standard define dams with a wall height > 15 m as large. The total global population of large dams with a wall height > 15 m is 45,000. There are 300 dams in the world of monumental scale; these "major dams" meet one of three criteria on height (>150 m), dam volume (>15 million $m^3$), or reservoir storage (>25 $km^3$) (Nilsson et al., 2005).

From this population of large dams, our reference class drew a representative sample of 245 dams (including 26 major dams) built between 1934 to 2007 on five continents in 65 different countries—the largest and most reliable data of its kind. The portfolio is worth USD 353 billion. All large dams for which valid and reliable cost and schedule data could be found were included in the sample. Of the 245 large dams, 186 were hydropower projects (including 25 major dams) and the remaining 59 were irrigation, flood control, or water supply dams. While we are primarily interested in the performance of large dam projects with a hydropower component, we also included non-hydropower dam projects in our reference class to test whether project types significantly differ in cost and schedule overruns or not. Figure 1 presents an overview of the sample by regional location, wall height, project type, vintage, and actual project cost.

**Figure 1: Sample distribution of 245 large dams (1934-2007), across five continents, worth USD 353B (2010 prices)**

| Location | Size - Dam Wall Height (m) |
|---|---|
| 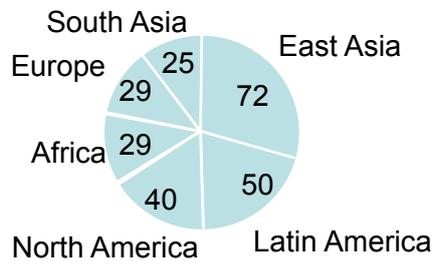 | 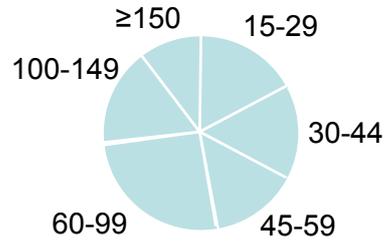 |

Location pie: South Asia 25, Europe 29, East Asia 72, Africa 29, North America 40, Latin America 50.

Size - Dam Wall Height (m) pie: ≥150, 15-29, 100-149, 30-44, 60-99, 45-59.

| Project Type | Project Vintage |
|---|---|
| 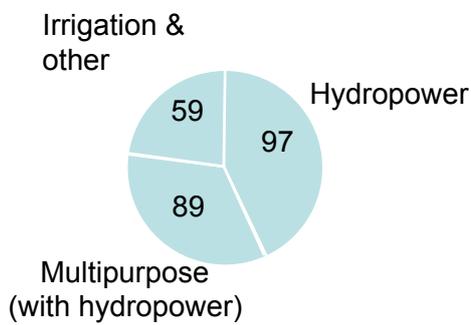 | 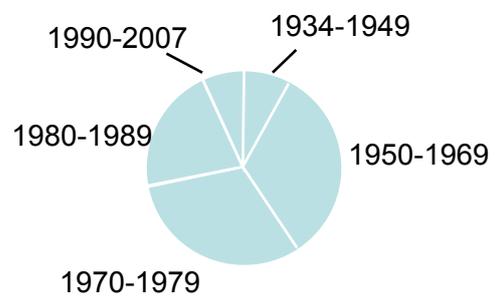 |

Project Type pie: Irrigation & other 59, Hydropower 97, Multipurpose (with hydropower) 89.

Project Vintage pie: 1990-2007, 1934-1949, 1980-1989, 1950-1969, 1970-1979.

**Actual Project Cost** (2010 USD millions), percent

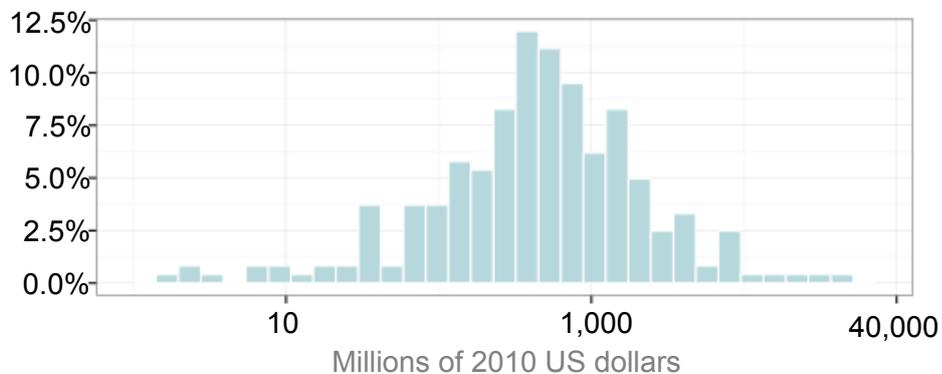

Millions of 2010 US dollars

The empirical strategy of this paper relied on documentary evidence on estimated versus actual costs of dams. Primary documents were collected from *ex ante* planning and *ex post* evaluation documents of the:

1. Asian Development Bank;

2. World Bank, also see World Bank (1996) and Bacon and Besant-Jones (1998);

3. World Commission of Dams (WCD), also see WCD (2000)[2];

4. U.S. Corps of Engineers;

5. Tennessee Valley Authority;

6. U.S. Bureau of Reclamation, also see Hufschmidt and Gerin (1970)[3] and Merewitz (1973) on the U.S. water-resource construction agencies.

The procedures applied to the cost and schedule data here are consistent with the gold standard applied in the field—more detailed methodological considerations can be found in Flyvbjerg et al. (2002); Federal Transit Administration (2003); Pickrell (1989, 1992); World Bank (1996) and Bacon and Besant-Jones (1998) with which our data are consistent. All costs are total project costs comprising the following elements: right-of-way acquisition and resettlement; design engineering and project management services; construction of all civil works and facilities; equipment purchases. Actual out-turn costs are defined as real, accounted construction costs determined at the time of project completion. Estimated costs are defined as budgeted, or forecasted, construction costs at the time of decision to build. The year of the date of the decision to build a project is the base year of prices in which all estimated and actual constant costs have been expressed in real (i.e. with the effects of inflation removed) local currency terms of the country in which the project is located. We exclude from our calculations debt payments, any *ex* *post* environmental remedial works, and opportunity cost of submerging land to form reservoirs. This makes comparison of estimated and actual costs of a specific project a like-for-like comparison.

*2.3. Analyses*

We investigated the magnitude and frequency of cost and schedule forecast (in)accuracies with a combination of simple statistical (parametric and non-parametric) tests and by fitting more sophisticated multilevel regression models sometimes termed Hierarchical Linear Models (HLM).

Multilevel or hierarchically structured data are the norm in the social, medical, or biological sciences. Rasbash et al. (2009, p. 1) explain: "For example, school education provides a clear case of a system in which individuals are subject to the influences of grouping. Pupils or students learn in classes; classes are taught within schools; and schools may be administered within local authorities or school boards. The units in such a system lie at four different levels of a hierarchy. A typical multilevel model of this system would assign pupils to level 1, classes to level 2, schools to level 3 and authorities or boards to level 4. Units at one level are recognized as being grouped, or nested, within units at the next higher level. Such a hierarchy is often described in terms of clusters of level 1 units within each level 2 unit etc. and the term clustered population is used." Important for a hierarchical linear model is that the dependent variable is at the lowest level of the nested structure. Multilevel models are necessary for research designs where data for observations are organized at more than one level (i.e., nested data) (Gelman and Hill, 2007). Failing to use multilevel models in such instances would result in spurious results (Rasbach et al., 2009).

With respect to our data on dams, projects are nested in the countries of their domicile. Like test scores of pupils from the same school tend exhibit within-school correlation, similarly outcomes of dam projects may exhibit within-country correlation that needs to be properly modeled using a multilevel model. We took this into account by modeling country as a random effect in a mixed effects multilevel model. The models were made parsimonious by using stepwise variable selection.

### 3. Results and interpretation

Our second step was to establish an empirical distribution for the cost forecast errors of large dams. We collected data on 36 possible explanatory variables, listed in Table 1, for the 245 large dams in our reference class.

**Table 1. Variables and characteristics used in multilevel regressions on construction cost overrun and schedule slip**

**Project-specific variables**
    **Project features**
        Hydropower or non-hydropower large dam project (dummy variable)
        New power station or station extension (dummy variable)
    **Size**
        Generator unit capacity (MW)
        Total project generation capacity (MW)
        Dam height for new hydropower station (meters)
        Hydraulic head for new hydropower station (meters)*
        Reservoir area created by project (hectares)*
        Length of tunnels (kilometers)*
    **Cost**
        Estimated project cost (constant local currency converted to 2010 USD MM)
        Actual project cost (constant local currency converted to 2010 USD MM)
        Cumulative inflation contingency (percentage)

**Time**
    Year of final decision to build
    Estimated implementation schedule (months)
    Year of start of full commercial operation
    Actual implementation schedule (months)

**Procurement**
    Estimated project foreign exchange costs as a proportion of estimated total project costs (percentage)
    Competitiveness of procurement process, international competitive bidding amount as a proportion of estimated total project costs (percentage)*
    Main contractor is from the host country (dummy variable)

**Country variables**
    Country (second level to control for within country correlation)
    Political regime of host country is a democracy (dummy variable)
    GDP of host country (current U.S. dollars)
    Per capita income of host country in year of loan approval (constant USD)
    Average actual cost growth rate in host country over the implementation period–the GDP deflator (percentage)
    MUV Index of actual average cost growth rate for imported project components between year of loan approval and year of project completion
    Long-term inflation rate of the host country (percentage)
    Actual average exchange rate depreciation or appreciation between year of formal-decision-to-build and year of full commercial operation (percentage)
    South Asian projects (dummy variable)
    North American projects (dummy variable)

\*   Denotes variables with a large number of missing values not used for regression analysis

*3.1 Preliminary statistical analysis of cost performance*

With respect to cost overruns, we make the following observations:

1. Three out of every four large dams suffered a cost overrun in constant local currency terms.

2. Actual costs were on average 96% higher than estimated costs; the median was 27% (*IQR* 0.86). The evidence is overwhelming that costs are systematically biased towards underestimation (Mann-Whitney-Wilcoxon $U = 29646$, $p < 0.01$); the magnitude of cost underestimation (i.e. cost overrun) is larger than the error of cost overestimation ($p <$ 0.01). The skew is towards adverse outcomes (i.e. going over budget).

3. Graphing the dams' cost overruns reveals a fat tail as shown in Figure 2; the actual costs more than double for 2 out of every 10 large dams and more than triple for 1 out of every 10 dams. The fat tail suggests that planners have difficulty in computing probabilities of events that happen far into the future (Taleb, [2007] 2010, p. 284).

**Figure 2. Density trace of Actual/Estimated cost (i.e. costs overruns) in constant local currency terms with the median and mean (N = 245)**

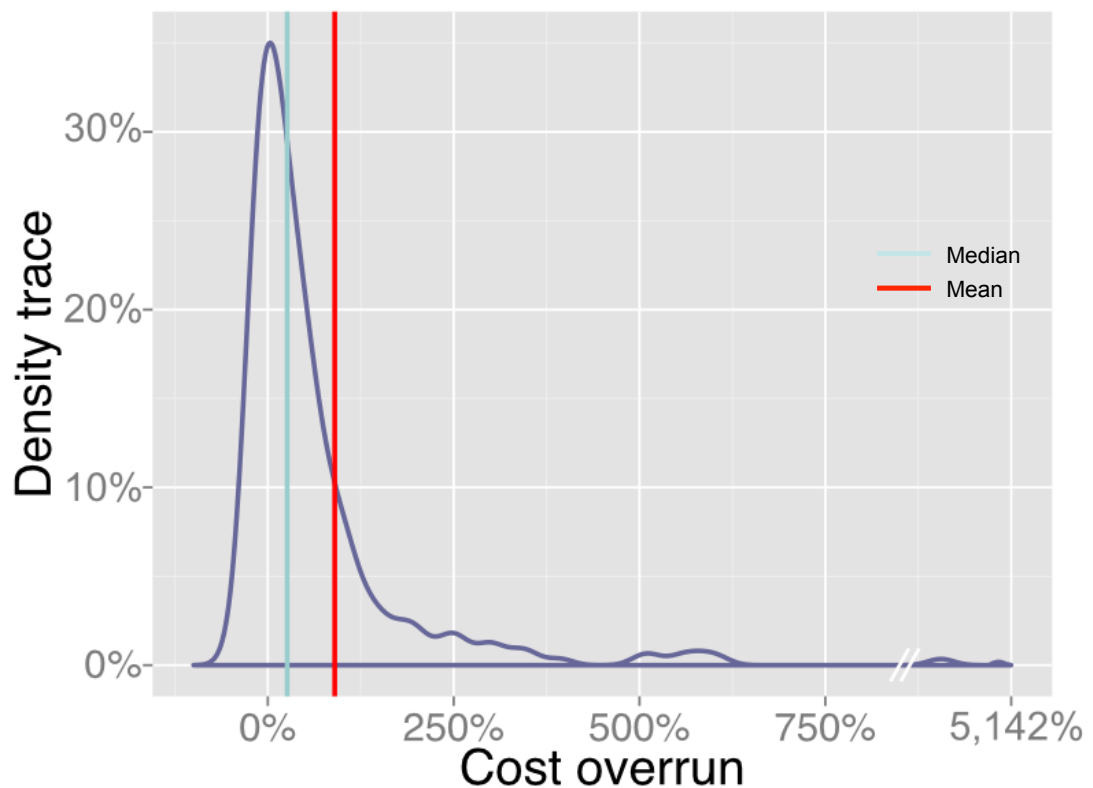

4. Large dams built in every region of the world suffer systematic cost overruns. The mean forecasting error is significantly above zero for every region. Figure 3 shows the geographical spread and cost over- runs of large dams in our reference class. Large dams built in North

America ($n$ = 40) have considerably lower cost overrun ($M$ = 11%) than large dams built elsewhere ($M$ = 104%). Although after controlling for other covariates such as project scale in a multilevel model, reported below, the differences among regions are not significant. We noted, 3 out of 4 dams in our reference class had a North American firm advising on the engineering and economic forecasts. Consistent with anchoring theories in psychology, we conjecture that an overreliance on the North American experience with large dams may bias cost estimates downwards in rest of the world. Experts may be "anchoring" their forecasts in familiar cases from North America and applying insufficient "adjustments" (Flyvbjerg et al., 2009; Tversky and Kahneman, 1974), for example to adequately reflect the risk of a local currency depreciation or the quality of local project management teams. Instead of optimistically hoping to replicate the North American cost performance, policymakers elsewhere ought to consider the global distributional information about costs of large dams.

**Figure 3. Location of large dams in the sample and cost overruns by**

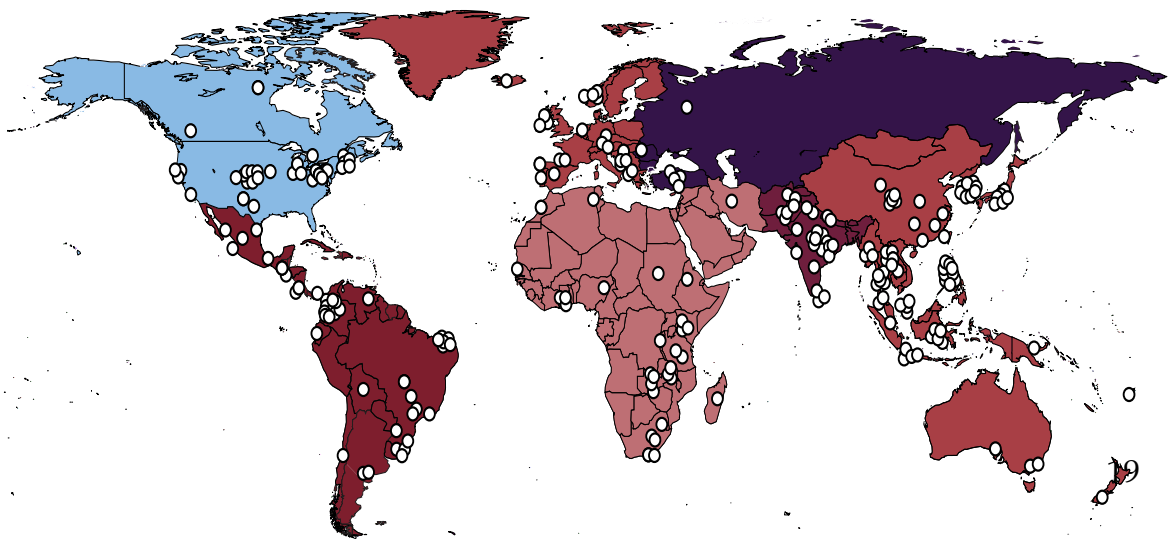

5. The typical forecasted benefit-to-cost ratio was 1.4. In other words, planners expected the net present benefits to exceed the net present costs by about 40%. Nearly half the dams suffered a cost overrun ratio of 1.4 or greater breaching this threshold after which the asset can be considered stranded—i.e. its upfront sunk costs are unlikely to be recovered. This is assuming, of course, that the benefits did not also fall short of targets, even though there is strong evidence that actual benefits of dams are also likely to fall short of targets (WCD, 2000; McCully, 2001; Scudder, 2005)[4].

6. We tested whether forecasting errors differ by project type (e.g., hydropower, irrigation, or multipurpose dam) or wall type (earthfill, rockfill, concrete arch etc.). Pairwise comparisons of percentage mean cost overrun and standard deviations as well as non-parametric Mann-Whitney tests for each of the parameters show no statistically significant differences. We conclude that irrespective of project or wall type, the probability distribution from our broader reference class of 245 dams applies as in Figure 2.

7. We analyzed whether cost estimates have become more accurate over time. Statistical analysis suggests that irrespective of the year or decade in which a dam is built there are no significant differences in forecasting errors ($F = 0.57$, $p = 0.78$). Similarly, there is no linear trend in-dicating improvement or deterioration of forecasting errors ($F = 0.54$, $p = 0.46$) as also suggested by Figure 4. There is little learning from past mistakes. By the same token, forecasts of costs of dams being made today are likely to be as wrong as they were between 1934-2007.

**Figure 4. Inaccuracy of cost estimates (local currencies, constant prices) for large dams over time (N=245), 1934-2007**

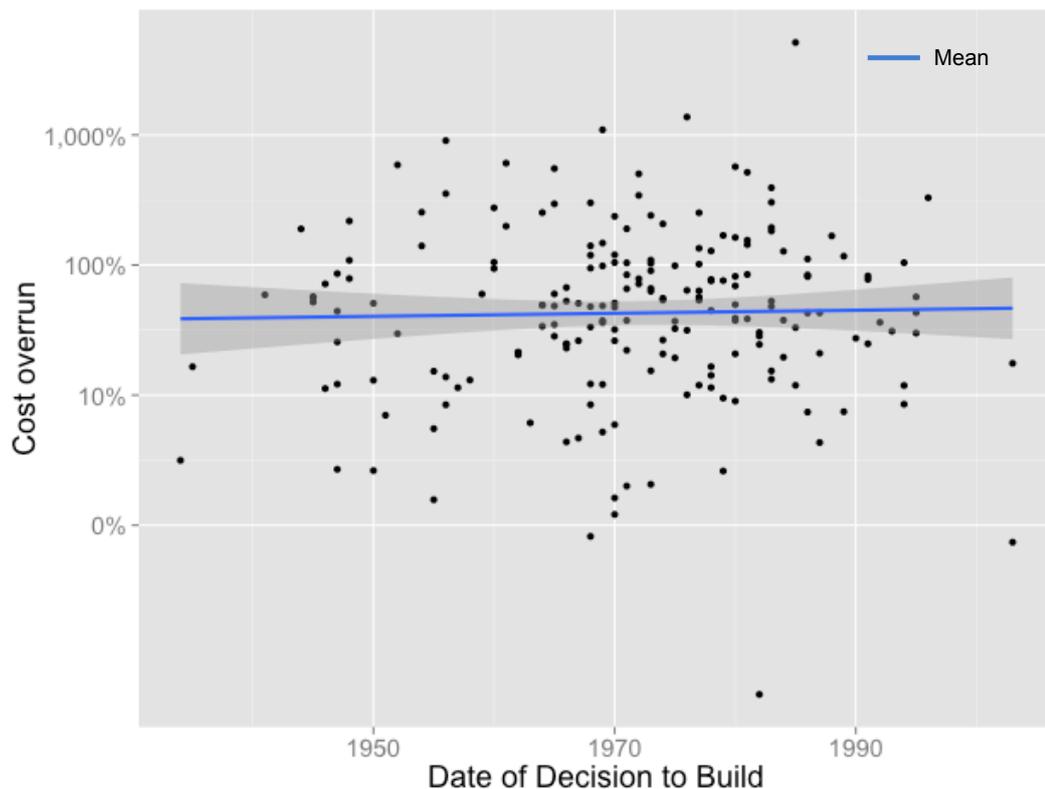

We also explored the absolute costs of large hydropower dams (N = 186). A large hydropower dam on average costs 1,800 million in 2010 USD with an average installed capacity of 630 MW. One MW installed capacity on average costs 2.8 million in 2010 USD. A preliminary univariate analysis, which makes no attempts to take into account any covariates, shows that increase in the scale of a dam, e.g., measured as height of the dam wall, increases the ab-solute investment required exponentially, e.g. a 100m high dam wall is four times more costly than a 50m wall ($R^2 = 0.27$, $F = 92.5$, $p < 0.01$). An even stronger relationship can be seen between installed capacity MW and actual costs ($R^2 = 0.70$, $F = 461.1$, $p < 0.01$).

Furthermore, the rate of cost overrun outliers increases with increase in dam size either measured in installed hydropower generation ($r = 0.24$, $p = 0.01$) or wall height ($r = 0.13$, $p = 0.05$). Since there is a significant correlation between dam height and hydropower installed capacity ($r = 0.47$, $p < 0.01$), evidence suggests that larger scale in general is prone to outlying cost overruns. We further investigate the effects of scale on cost overruns by fitting multilevel models (model 1 and 2) reported below.

*3.2 Preliminary statistical analysis of schedule performance*

Not only are large dams costly and prone to systematic and severe budget overruns, they also take a long time to build. Large dams on average take 8.6 years. With respect to schedule slippage, we make the following observations:

8. Eight out of every 10 large dams suffered a schedule overrun

9. Actual implementation schedule was on average 44% (or 2.3 years) higher than the estimate with a median of 27% (or 1.7 years) as shown in Figure 5. Like cost overruns, the evidence is overwhelming that implementation schedules are systematically biased towards underestimation (Mann-Whitney-Wilcoxon $U = 29161$, $p < 0.01$); the magni-tude of schedule underestimation (i.e. schedule slippage) is larger than the error of schedule overestimation ($p < 0.01$).

10. Graphing the dams' schedule overruns also reveals a fat tail as shown in Figure 5, albeit not as fat as the tail of cost overruns. Costs are at a higher risk of spiraling out of control than schedules.

**Figure 5. Density trace of schedule slippage (N = 239) with the median and mean**

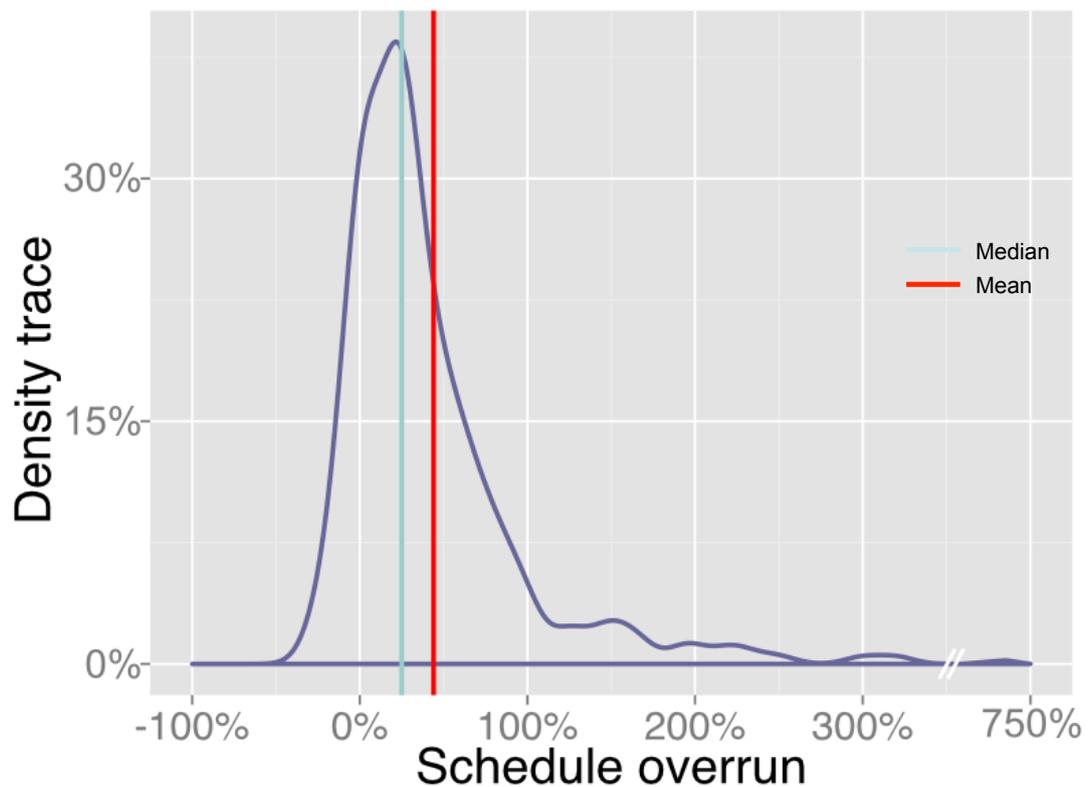

11. There is less variation in schedule overruns across regions than cost overruns. Large dams built everywhere take significantly longer than planners forecast. North America with a 27% mean schedule overrun is the best performer. A non-parametric comparison using a Wilcoxon test ($p = 0.01$) suggests that projects in South Asia have significantly greater schedule overruns ($M = 83\%$) than rest of the world taken as a whole ($M$ = 42%). We investigate this further with a multilevel model below (model 3).

12. There is no evidence for schedule estimates to have improved over time.

We tested whether implementation schedules and project scale are related. A preliminary univariate analysis, which makes no attempts to take into account any covariates, shows that increase in the scale of a dam, e.g., measured as estimated cost of construction, increases the absolute actual implementation schedule required exponentially ($R^2$ = 0.13, $F$ = 36.4, $p$ < 0.01). Large scale is intimately linked with the long-term (see model 2 below). The actual implementation schedule, reported here, does not take into the account lengthy lead times in preparing the projects. Dams require extensive technical and economic feasibility analysis, social and environmental impact studies, and political negotiations. The actual implementation cycles are far longer than the average of about 8.6 years, as shown in our data, that it takes to build a dam. These lengthy implementation schedules schedule suggest that the benefits of large dams (even assuming that large dam generate benefits as forecasted) do not come 'online' quickly enough. The temporal mismatch between when users need specific benefits and when these benefits come online is not to be downplayed (Ansar et al., 2012). Alternative investments that can bridge needs quickly, without tremendous time lags, are preferable to investments with a long lead-time and hence duration risk (Luehrman, 1998; Copeland and Tufano, 2004).

*3.3 Multilevel regression analysis of cost and schedule performance*

Means, standard deviations, and correlations of the variables used in the multilevel regressions are shown in Table 2.

**Table 2. Descriptive Statistics and Correlations (N = 245)**

| Variable | Mean | S.D. | 1 | 2 | 3 | 4 | 5 | 6 | 7 | 8 | 9 |
|---|---|---|---|---|---|---|---|---|---|---|---|
| 1. Cost Overrun[a] | 2.0 | 3.6 | | | | | | | | | |
| 2. Schedule slippage[a] | 1.5 | 0.7 | 0.17** | | | | | | | | |
| 3. Estimated schedule (months)[b] | 73.1 | 33.8 | -0.16* | 0.23** | | | | | | | |
| 4. Actual schedule (months)[b] | 102.7 | 55.7 | -0.27** | -0.43** | 0.76** | | | | | | |
| 5. Year - decision to build | 1971.1 | 13.2 | -0.02 | 0.05 | -0.21** | -0.25** | | | | | |
| 6. Year - completion | 1979.6 | 12.7 | -0.14* | -0.10 | 0.03 | 0.08 | 0.94** | | | | |
| 7. Project type dummy | 0.8 | 0.4 | -0.14* | 0.08 | 0.10 | 0.02 | -0.02 | -0.02 | | | |
| 8. Democracy dummy | 0.4 | 0.5 | 0.00 | -0.14* | 0.16* | 0.20** | -0.45** | -0.38** | 0.00 | | |
| 9. Estimated cost (US$ MM 2010 constant)[b] | 699.6 | 1215.5 | -0.03 | 0.09 | 0.48** | 0.37** | 0.02 | 0.13* | 0.37** | -0.04 | |
| 10. Actual cost (US$ MM 2010 constant)[b] | 1462.2 | 4032.5 | -0.38** | 0.02 | 0.50** | 0.43** | 0.02 | 0.17** | 0.38** | -0.03 | 0.93** |
| 11. Height of dam wall (m)[c] | 77.3 | 51.6 | -0.10 | 0.10 | 0.26** | 0.17** | 0.10 | 0.16* | 0.34** | -0.03 | 0.51** |
| 12. Installed hydropower capacity (MW)[b] | 487.0 | 1255.3 | -0.16* | 0.19** | 0.22** | 0.08 | 0.13* | 0.16* | 0.69** | -0.14* | 0.59** |
| 13. Length of dam wall (m)[b] | 1364.1 | 2061.9 | -0.12 | -0.07 | 0.25** | 0.30** | -0.19** | -0.08 | -0.07 | 0.08 | 0.37** |
| 14. Tunnel length (m)[b] | 3500.0 | 7869.5 | 0.13 | -0.12 | -0.04 | 0.16 | -0.06 | -0.01 | -0.23 | 0.05 | 0.11 |
| 15. Manufactures Unit Value Index CAGR[d] | 6.0 | 5.4 | -0.01 | -0.03 | -0.25** | -0.18** | -0.12 | -0.18** | 0.08 | -0.08 | -0.13 |
| 16. GDP (nominal US$ B)[b] | 1221.1 | 253.4 | -0.05 | 0.25** | 0.36** | 0.17* | 0.29** | 0.37** | -0.13 | 0.13 | 0.19* |
| 17. Per capita income (2000 constant US$)[b] | 4132.8 | 5198.6 | 0.23** | 0.15* | 0.11 | 0.01 | -0.37** | -0.40** | -0.07 | 0.48** | -0.07 |
| 18. Long-term inflation (%)[b] | 17% | 0.2 | -0.29** | 0.04 | -0.09 | -0.11 | 0.22** | 0.19** | 0.24** | -0.37** | 0.13* |
| 19. Forex depreciation (%)[e] | 18% | 70.3 | -0.30** | -0.04 | 0.03 | 0.00 | 0.29** | 0.29** | 0.16* | -0.20** | 0.21** |
| 20. South Asia dummy | 0.1 | 0.3 | -0.25** | -0.18** | 0.17** | 0.26** | -0.04 | 0.07 | -0.06 | 0.20** | 0.11 |
| 21. North America dummy | 0.2 | 0.4 | 0.28** | 0.06 | 0.21** | 0.13* | -0.57** | -0.55** | -0.09 | 0.52** | 0.06 |
| Variable | 10 | 11 | 12 | 13 | 14 | 15 | 16 | 17 | 18 | 19 | 20 |
| 11. Height of dam wall (m) | 0.51** | | | | | | | | | | |
| 12. Installed hydropower capacity (MW) | 0.60** | 0.47** | | | | | | | | | |
| 13. Length of dam wall (m) | 0.38** | 0.03 | 0.13 | | | | | | | | |
| 14. Tunnel length (m) | -0.01 | 0.05 | -0.22 | -0.18 | | | | | | | |
| 15. Manufactures Unit Value Index CAGR | -0.12 | -0.08 | -0.02 | 0.02 | -0.02 | | | | | | |
| 16. GDP (nominal USD) | 0.19* | 0.10 | 0.09 | 0.04 | -0.29 | -0.31** | | | | | |
| 17. Per capita income (2000 constant USD) | -0.14* | -0.08 | -0.11 | -0.02 | -0.09 | -0.01 | 0.29** | | | | |
| 18. Long-term inflation (%) | 0.22** | 0.06 | 0.33** | 0.07 | -0.41* | 0.15* | -0.03 | -0.24** | | | |
| 19. Forex | 0.29** | 0.09 | 0.29** | -0.02 | -0.37* | -0.16* | 0.00 | -0.26** | 0.64** | | |
| 20. South Asia dummy | 0.19** | 0.08 | -0.03 | 0.20** | NA | -0.09 | -0.01 | -0.46** | -0.10 | 0.11 | |
| 21. North America dummy | -0.03 | -0.10 | -0.16* | 0.19** | NA | -0.18** | 0.33** | 0.60** | -0.44** | -0.31** | -0.15* |

[a] One over (1/$x$) transformed; [b] Log transformed; [c] Sq. rt. ($\sqrt{x}$); [d] Cb rt. ($\sqrt[3]{x}$); [e] $x$^0.25 transformed to remove excess skewness for regression analysis and to calculate correlations. Please bear transformations in mind in interpreting correlation coefficient signs.
** $p < .01$; * $p < .05$

We fitted multilevel regression models with projects nested in country as a second level to incorporate within-country correlation. The models were fitted using the "lme" procedure in the "nlme" package in R software. This function fits a linear mixed-effects model in the formulation described in Laird and Ware (1982) but allowing for nested random effects. The within-group errors are allowed to be correlated and/or have unequal variances. We found it necessary to transform variables to remove excessive skewness as noted in Table 2. Using stepwise variable selection, we are not only able to fit explanatory models for cost and overruns and estimated duration but also practicably parsimonious models for predicting them.

Table 3 summarizes the results from multilevel model examining predictors of cost overruns (model 1). Model 1 identifies the estimated implementation schedule and the long-term inflation rate in the country in which the project is built as highly significant variables. An increase in estimated duration of one year contributes to an increase in cost overrun of approx. 5%-6% depending on the country whilst holding the inflation rate constant (see figure in Appendix 1). Note that a R-squared measure, which is customary to report for single-level regressions as explained proportion of variance, cannot be applied to multilevel models (Recchia, 2010)[5]. The usual diagnostics, based upon the model residuals, were satisfactory.

**Table 3. Model 1 - Significant variables for cost accuracy for large dam projects (constant local currency)**

| Variable | Regression coefficient | Standard error | t-stat | 2-tailed significance |
|---|---|---|---|---|
| Intercept | 1.402 | 0.185 | 7.560 | 0.000 |
| Log estimated duration (months) | -0.100 | 0.041 | -2.424 | 0.016 |
| Log of country's long-term inflation rate (%) | -0.085 | 0.029 | -2.930 | 0.005 |

*Note*: Dependent variable is the estimated/actual cost ratio (i.e. 1/*x* of the cost overrun to remove excessive skewness), based on 239 observations. Since the dependent variable in Model 1 is the inverse of the cost overrun a negative sign on the coefficients of both significant variables suggests that an increase in the estimated duration or long-term inflation rate increases the cost overrun.

The first finding in Model 1 is that the larger the estimated implementation schedule the higher the cost overrun ($p = 0.016$), all other things being equal, and is particularly noteworthy for two reasons.

> First, Model 1 suggests that planners' forecasting skills decay the longer in the future they are asked to project the risks facing a large dam. Material information about risks, for example related to geology, prices of imports, exchange rates, wages, interest rates, sovereign debt, environment, only reveals in future shaping episode to which decision-makers are "blind" *ex ante* (Flyvbjerg and Budzier, 2011). We discuss some qualitative case examples to illustrate this statistical result and its broader implications in the next section.

> Second, preliminary analysis had suggested that estimated implementation schedules depend on the scale of a planned investment—i.e. bigger projects take longer to build. Support of this preliminary result was found by fitting a multilevel model (model 2) that examines the predictors of estimated implementation schedule. Model 2 shows that height ($p = 0.02$), the installed capacity (MW) ($p = 0.02$), and length ($p = 0.04$) of the dam wall are significant variables associated with the estimated implementation schedule. The effect of these covariates can be seen from the coefficients in Table 4: a greater height, installed capacity, or length contributes to a longer implementation schedules. We interpret Model 2 as follows. Estimated implementation schedule not only acts as a temporal variable but also as a surrogate for scalar variables such as wall height (which is also highly correlated with installed capacity). The larger the dam, the longer the estimated implementation schedule, and the higher the cost overrun.

Taken together, the multilevel models for cost overruns and estimated schedule suggest that longer time horizons and increasing scale are underlying causes of risk in investments in large hydropower dam projects.

**Table 4. Model 2 - Significant variables for estimated construction schedule for large dam projects (months)**

| Variable | Regression coefficient | Standard error | t-stat | 2-tailed significance |
|---|---|---|---|---|
| Intercept | 3.444 | 0.197 | 17.464 | 0.000 |
| Sq rt of dam wall height (m) | 0.029 | 0.012 | 2.414 | 0.017 |
| Log of dam wall length (m) | 0.058 | 0.027 | 2.153 | 0.033 |
| Log of hydropower installed capacity (MW) | 0.016 | 0.007 | 2.141 | 0.034 |

*Note*: Dependent variable is log of the estimated construction schedule, based on 239 observations.

The second finding in Model 1 is that higher the long-term inflation rate of the host country the higher the cost overrun suffered by a dam ($p = 0.02$). The long-term inflation rate was calculated by fitting a linear model to the log of the time series of the GDP deflator index of each country. The slope of this fitted line can be interpreted as the annual average growth rate of the log inflation for each country. This slope is a different constant for each country with some countries such as Brazil with a considerably higher long-term inflation rate, and hence greater propensity to cost overruns, than China or the United States. Moreover, this slope is stable in the short-run (it takes years of high or low inflation to change this slope) and hence our estimate can be assumed to be a reliable predictor. Recall that the cost overrun is being measured in constant terms (i.e. with the effects of inflation removed); yet Model 1 suggests that the inflation trajectory of a country, which we interpret as a surrogate of the overall macroeconomic management, is an important risk when making durable investments. The multilevel model thus suggests that once country specific factors have been taken into account the factor that drives cost overrun is the planning horizon.

Finally, we fit a multilevel model (Model 3) to examine predictors of schedule overruns. Model 3 identifies the following significant variables: whether or not a country is a democracy; the per capita income of the country in 2000 constant USD in the year of the decision to build; the planned installed capacity (MW); and planned length of the dam wall (meters). Avid dam building countries in South Asia, at various stages of democratic maturity, have also one of the poorest schedule performances in building dams. We controlled for this fact by including a dummy variable for South Asia in the model as a covariate with an interaction effect with the democracy dummy. Democracy in South Asia is significant in explaining schedule overruns. The South Asia dummy, however, does not come out to be significant. The effect of these covariates and the interaction effect can be seen in Table 5.

**Table 5. Model 3 - Significant variables for schedule slippage for large dam projects**

| Variable | Regression coefficient | Standard error | *t*-stat | 2-tailed significance |
|---|---|---|---|---|
| Intercept | 0.405 | 0.163 | 2.483 | 0.014 |
| Democracy dummy* | -0.134 | 0.055 | -2.439 | 0.016 |
| Log of country's per capita income in year of decision to build (constant USD) | 0.065 | 0.019 | 3.334 | 0.001 |
| Log of dam wall length (m) | -0.027 | 0.013 | -2.081 | 0.039 |
| Log of hydropower installed capacity (MW) | 0.018 | 0.006 | 3.207 | 0.002 |
| South Asia dummy | 0.211 | 0.113 | 1.874 | 0.066 |
| Democracy in South Asia interaction effect | -0.239 | 0.113 | -2.114 | 0.036 |

*Dummy based on the Polity2 variably of Polity IV regime index. Score of +10 to +6 = democracy; score of +5 to -10 = autocracy.

*Note*: Dependent variable is 1/*x* of the actual/estimated schedule ratio, based on 239 observations[6].

First, democracies' forecasts about implementation schedules of large dams are systematically more optimistic than autocracies even after controlling for systematically higher schedule overruns in India and Pakistan. The size of the coefficient is large suggesting that political process has profound impact on the schedule slippage. We tested whether democracies take longer than autocracies to build large dams by fitting a model to explain the actual implementation schedule (Model 4). Model 4, summarized in Table 6, shows that effects of political regime on the actual schedule are not significant. In other words, while democracies do not take longer to build large dams than autocracies in absolute terms, democracies appear to be more optimistic. Given its vast scope, we defer a further investigation of this important result to a future enquiry. We note, however, that theories of delusion and deception in the planning of large infrastructure projects (Flyvbjerg et al., 2009) would interpret this as evidence of *ex ante* political intent among democratically elected politicians to present a rosier picture about large dams than they know the case to be.

Second, countries with a higher per capita income in constant 2000 USD in the year of decision to build tend to have lower schedule overruns than countries with lower per capita income. We concur with the interpretation of Bacon and Besant-Jones (1998, p. 325) that "the best available proxy for most countries is [the] country-per-capita income…[for] the general level of economic support that a country can provide for the construction of complex facilities". This result suggests that developing countries in particular, despite seemingly the most in need of complex facilities such as large dams, ought to stay away from bites bigger than they can chew.

**Table 6. Model 4 - Significant variables for estimated construction schedule for large dam projects (months)**

| Variable | Regression coefficient | Standard error | *t*-stat | 2-tailed significance |
|---|---|---|---|---|
| Intercept | -17.712 | 6.401 | -2.767 | 0.007 |
| Log of dam wall length (m) | 0.105 | 0.029 | 3.567 | 0.001 |
| Year of actual project completion | 0.011 | 0.003 | 3.358 | 0.001 |

*Note*: Dependent variable is log of the actual construction schedule, based on 239 observations.

Third, the evidence appears to be contradictory with respect to scale. While a greater dam wall length contributes to a higher schedule overrun, a higher MW installed capacity has the opposite effect. Model 3 in Table 5 shows that the size of coefficients for the two significant variables related to physical scale—i.e. Log of dam wall length (m) and Log of hydropower installed capacity (MW)—is approximately the same but with the opposite sign[6].

In attempting to interpret this result our conjecture is as follows. Dam walls are bespoke constructions tied to the geological and other site-specific characteristics. In contrast, installed capacity is manufactured off-site in a modular fashion. For example, the 690 MW installed capacity of the recently completed Kárahnjúkar project in Iceland was delivered with six generating units of identical design (6 X 115 MW). We propose that project components that require onsite construction, e.g. dam wall, are more prone to schedule errors than components manufactured off-site, e.g. generation turbines. Project designs that seek to reduce the bespoke and onsite components in favor of greater modular and manufactured components may reduce schedule uncertainty.

This conjecture is supported by Model 4 in Table 6, which shows that the actual construction schedule, in absolute terms, is significantly increased with an increase in the length of dam wall. In contrast, MW installed capacity does not have an effect on the absolute actual construction schedule suggesting that construction schedules are more sensitive to on-site construction than to components manufactured in factories. Note that lower installed capacity does not necessarily equate with a smaller dam. For example, it is not rare for a large multipurpose dam to have a low MW installed capacity when, for instance, the dam is primarily being used for irrigation or flood management purposes.

## 4. Qualitative case examples and policy propositions

The statistical results reported in the preceding sections show that cost and schedule estimates of large dams are severely and systematically biased below their actual values. While it is beyond the scope of this paper to discuss wider theoretical implications, the evidence presented here is consistent with previous findings that point to twin problems that cause adverse outcomes in the planning and construction of large and complex facilities such as large hydropower dams: 1) biases inherent in human judgment (delusion) and 2) misaligned principal-agent relationships or political incentives (deception) that underlie systematic forecasting errors. In context of large dams, we argue that large scale and longer planning time horizons exacerbate the impact of these twin problems. We now present a few qualitative examples of risks large dams typically face to illustrate the statistical results reported above. We jointly draw on the statistical analyses and qualitative analyses to distill propositions of immediate relevance to policy.

Globally, experts' optimism about several risk factors contribute to cost overruns in large dams. For example, the planning documents for the Itumbiara hydroelectric project in Brazil recognized that the site chosen for the project was geologically unfavorable. The plan optimistically declared, "the cost estimates provide ample physical contingencies [20% of base cost] to provide for the removal of larger amounts [of compressible, weak, rock] if further investigations show the need" (World Bank, 1973). This weak geology ended up costing +96% of the base cost in real terms. Itumbiara's case is illustrative of a broader problem. Even though geological risks are anticipatable there is little planners can do to hedge against it. For example, exhaustive geological investigation for a large dam can cost as much as a third of the total cost (Hoek and Palmieri, 1998); at which point still remains a considerable chance of encountering unfavorable conditions that go undetected during the *ex ante* tests (Goel et al., 2012).

> *Policy proposition 1: Energy alternatives that rely on fewer site-specific characteristics such as unfavorable geology are preferable.*

Similarly, in the Chivor hydroelectric project in Colombia, the planning document was upbeat that there will be no changes in the exchange rate between the Colombian Peso and the U.S. dollar during the construction period (1970-1977) stating, "No allowance has been made for possible future fluctuations of the exchange rate. This approach is justified by recent experience in Colombia where the Government has been pursuing the enlightened policy of adjusting [policy] quickly to changing conditions in the economy" (World Bank, 1970). In fact, the Colombian currency depreciated nearly 90% against the U.S. dollar as shown in Figure 6.

**Figure 6. Depreciation of the Colombian Peso 1970-2010**

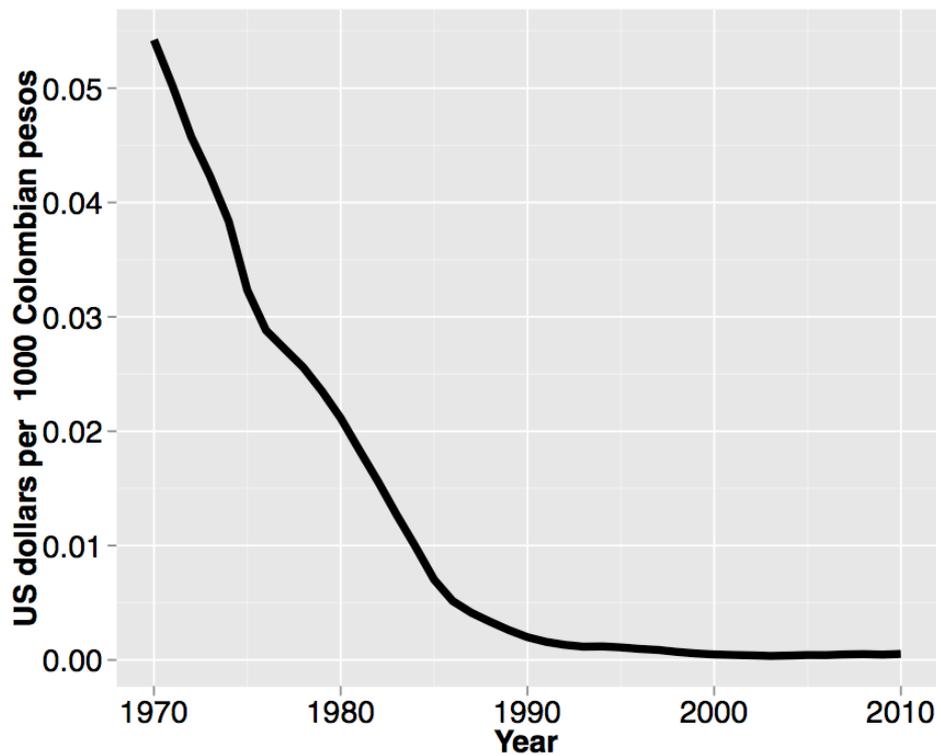

Since over half the project's costs covers imported inputs, this currency depreciation caused a 32% cost overrun in real Colombian Peso terms. Currency exposure arises when the inputs required to build a project are denominated in one currency but the outputs in another, or vice versa. The outputs of dams, such as electricity, are denominated in the local currency. Similarly, any increases in tax receipts a dam may enable for the host government also accrue in local currency. A large portion of inputs to build a dam, particularly in developing countries, however, constitute imports paid for in USD. Since the USD liabilities also have to eventually be paid in local currency, currency exposure consistently proves to be a fiscal hemorrhage for large projects.

> *Policy proposition 2: Energy alternatives that rely on fewer imports or match the currency of liabilities with the currency of future revenue are preferable.*

Although, following convention, our cost analysis excludes the effects of inflation, planners ought not to ignore the risks of "unanticipated inflation" (Pickrell, 1992, p. 164). Episodes of hyperinflation in Argentina, Brazil, Turkey, and Yugoslavia caused staggering nominal cost overruns, e.g.7-times initial budget for Brazil's Estreito dam (1965-1974), or 110-times initial budget for Yugoslavia's Visegrad dam (1985-1990), which had to be financed with additional debt. Effects of unanticipated inflation magnify the longer it takes to complete a project. For example, during the planning phase of Pakistan's Tarbela dam, it was assumed that inflation would not have a signification impact on the project's costs. The appraisal report wrote: "A general contingency of 7½% has been added in accordance with normal practice for works of this size and duration" (World Bank, 1968). The project, launched in 1968, was meant to start full commercial operation in 1976, but the opening was delayed until 1984. Actual cumulative inflation in Pakistan during 1968-1984 was 380%; the actual cost of the dam in nominal terms nearly four times the initial budget. In the case of Tarbela, unanticipated inflation was "a product of delays in a project's construction timetable and a higher-than expected inflation rate" (Pickrell, 1992, p. 164). For our reference class, 8 out of 10 large dams came in late with an average delay of 2.3 years. Moreover, forecasters expected the annual inflation rate to be 2.5% but it turned out to be 18.9% (averages for the entire sample). Large dams have a high propensity to face unanticipated inflation.

> *Policy proposition 3: The best insurance against creeping inflation is to reduce the implementation schedule to as short a horizon as possible. Energy alternatives that can be built sooner and with lower risk of schedule overruns, e.g. through modular design, are preferable.*

Large dams are typically financed from public borrowing. While our calculations exclude debt-servicing, cost overruns increase the stock of debt but also the recurring financing costs that can further escalate if interest rates go up. The optimistic risk assessments of the costs of large dams are consistent with "explosive growth of Third World debt" (Bulow and Rogoff, 1990; Mold, 2012). For example, the actual cost of Tarbela dam, the majority of which was borrowed from external sources, amounted to 23% of the increase in Pakistan's external public debt stock between 1968-1984; or 12% for Colombia's Chivor dam (1970-1977) as shown in Table 7.

**Table 7. Total stock of public net external debt (US$ current, MM)**

| Year | Colombia | Pakistan |
|---|---|---|
| 1968 | | 3,252.4 |
| 1970 | 1,296.6 | |
| 1977 | 2,699.6 | |
| 1984 | | 9,692.8 |
| Debt increase over the implementation schedule | 1,403.0 | 6,440.5 |
| Cost of mega-dam (US$ current MM) | Chivor dam | Tarbela dam |
| | 168.7 | 1,497.9 |
| Cost of dam as percentage of debt increase | | |
| | 12.0% | 23.0% |

These case examples reinforce the essential message of statistical results: bigger projects entail uncontrollable risks, which even when anticipatable cannot be adequately hedged. We do not directly negate the presence of economies of scale or learning curves—i.e. declining average cost per unit as output increases. Instead our argument is that any economies of scale embedded in large scale are being acquired for a disproportionately increased exposure to risk that can cause financial impairment. Companies and countries with insufficient capacity to absorb adverse outcomes of big bets gone awry often face financial ruin.

*Policy proposition 4: Energy alternatives that do not constitute a large proportion of the balance sheet of a country or a company are preferable.*

*Similarly, policymakers, particularly in countries at lower levels of economic development, ought to avoid highly leveraged alternatives investments denominated in a mix of currencies.*

## 5. Forecasting the actual costs and schedules using reference class forecasting (RCF)

As discussed in the methods section, the third step of the "outside view" or RCF techniques is to compare a specific venture with the reference class distribution, in order to establish the most likely outcome for the specific venture. Thus if systematic errors in the forecasts generated using the "inside view" of previous ventures are found, decision-makers should apply an uplift or downlift to the "inside view" forecast in order to generated a debiased "outside view" forecast. For example, empirical literature has established that rail projects suffer a cost overrun of 45% on average (Flyvbjerg, 2008; also see Table 8). The 50$^{th}$ percentile cost overrun for rail projects is 40% and the 80$^{th}$ percentile is 57%. Based on these findings, RCF techniques suggest that decision-makers ought to apply a 57% uplift to the initial estimated budget in order to obtain 80% certainty that the final cost of the project would stay within budget (Flyvbjerg 2008, p.16). If decision-makers were more risk tolerant then they could apply a 40% uplift to the initial estimated budget but then there will remain a 50% chance that the proposed project might exceed its budget.

In line with the RCF techniques, the third and final step of our investigation on dams was to derive a good predictor of cost and schedule overruns for proposed large dams based on the distributional information of the reference class. This predictor serves to "correct" the systematically biased *ex ante* cost and schedule estimates by adjusting them upwards by the average cost or schedule overrun (see Kahneman and Tverksy, 1979b; Flyvbjerg, 2006, 2008).

First, using traditional RCF (Flyvbjerg, 2006, 2008), we traced the empirical distribution of cost and schedule overruns of large dams. Second, we use multilevel Models 1 and 3, described above, for predicting cost and schedule overruns. Model 1 and 3 prove to be practicably parsimonious models for two reasons: First both models are fitted with variables known *ex ante*. Second, both models were successfully fitted with only a few significant variables making it practicable to collect the data needed to make a prediction. For example, Model 1 on cost overruns has only two significant variables—estimate schedule and the long-term inflation rate of the host country. Data on both these variables is readily available for any proposed large dam making it possible to predict the cost overrun before construction begins. We illustrate the usefulness of our predictive models with an example below.

With respect to cost overruns, using traditional RCF (Flyvbjerg, 2006, 2008), we find that if planners are willing to accept a 20% risk of a cost overrun, the uplift required for large dams is +99% (i.e. ~ double experts' estimates) as seen in Figure 7; and +176% including unanticipated inflation. If planners are willing to accept a 50-50 chance of a cost overrun, the uplift required is 26% (32% outside North America).

**Figure 7: Required uplift for large dam projects as function of the maximum acceptable level of risk for cost overrun, constant local currency terms (N = 245)**

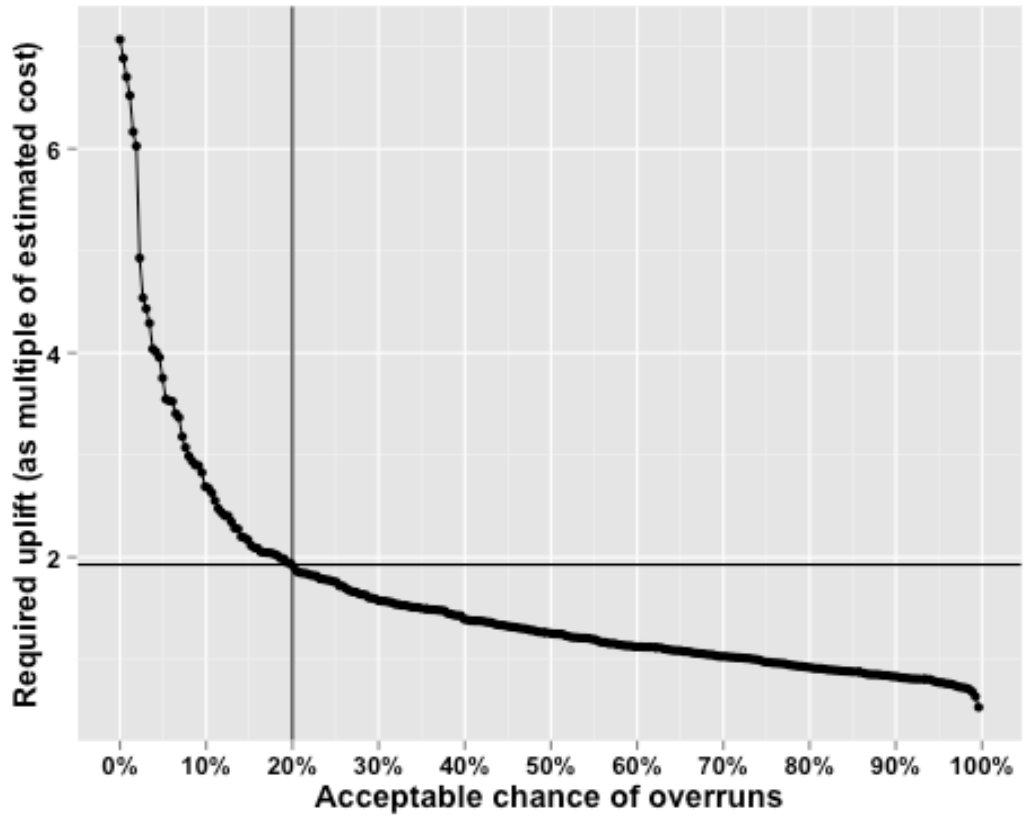

In terms of cost overruns, Figure 7 also illustrates that large dams are one of the riskiest asset classes for which valid and reliable are available. Compare, for example, Figure 7 with reference class forecasts previously conducted for rail, road, tunnel, or bridge projects (Flyvbjerg, 2006, 2008) also summarized in Table 8.

**Table 8. Comparing large dams with other infrastructure asset classes**

|  | Applicable capital expenditure optimism bias uplifts (constant prices) |
|---|---|

| Category | Types of projects | Mean cost overrun | 50th percentile | 80th percentile |
|---|---|---|---|---|
| Roads | Motorway<br>Trunk roads<br>Local roads<br>Bicycle facilities<br>Pedestrian facilities<br>Park and ride<br>Bus lane schemes<br>Guided buses | 20% | 15% | 32% |
| Rail | Metro<br>Light rail<br>Guided buses on tracks<br>Conventional rail<br>High speed rail | 45% | 40% | 57% |
| Fixed links | Bridges<br>Tunnels | 34% | 23% | 55% |
| Building projects | Stations<br>Terminal buildings | | 4 – 51%[a] | |
| Standard civil engineering | | | 3 – 44%[a] | |
| Non-standard civil engineering | | | 6 – 66%[a] | |
| Mining projects | | 14%[b] | | |
| Thermal power plants | | 6%[c] | | |
| Large dam projects | Large hydropower<br>Large Irrigation<br>Flood control<br>Multipurpose dams | **96%** | 26% | **99%** |
| Nuclear power plants | | 207%[d] | 109 – 281%[d] | |

[a] Based on Mott MacDonald (2002); [b]Based on Bertisen and Davis (2008); [c]Based on Bacon and Besant-Jones (1998, p.321), included for an approximate comparison purposes only, reference class probability distribution not available. dBased on Schlissel and Biewald (2008, p.8) review of the U.S. Congressional Budget Office (CBO) data from Energy Information Administration, Technical Report DOE/EIA-0485 (January 1, 1986).

Second, using our multilevel Model 1 we were able to derive predictions for cost overrun (in constant local currency) and schedule overrun respectively.

Experts estimate, for instance, that Pakistan's Diamer-Bhasha dam, whose construction began shortly after the 2010 floods, will cost PKR 894 billion (~USD12.7B in 2008 prices and exchange rates and about 9% of Pakistan's 2008 GDP) (WAPDA, 2011). The dam is forecasted to take ten years from 2011 and become operational in 2021. Using our first approach, the reference class forecast for cost overruns suggests that planners need to budget PKR 1,788B (USD25.4B) in real terms to obtain 80% certainty of not exceeding the budget. Including the effects of unanticipated inflation the required budget is PKR 2,467B (USD35.0B) or about 25% of Pakistan's 2008 GDP. A future sovereign default in Pakistan owing to this one mega-dam is not a remote possibility.

Using our second approach, our multilevel Model 1 predicts that given the 10 year estimated duration and a long-term inflation rate of about 8% the expected (average) cost overrun of a large dam in Pakistan will be 44% (PKR 1,288B or USD 18.3B). Combining the two methods, a conservative estimate for the cost overrun on the Diamer-Bhasha dam is 44% at which point there remains a 4 in 10 chance of the revised budget being exceeded. Note, however, that if a dam of dimensions similar to Diamer-Bhasha were being built in the US, Model 1 predicts that it would only suffer a cost overrun of 16%, which the much larger US economy could absorb without any lasting damage.

We applied a similar two-pronged forecast of schedule slippage. Using our first approach, the reference class forecast for schedule slippage suggests that planners for large dams around the world need to allow for a 66% schedule overrun to achieve 80% certainty that the project will be completed within the revised implementation schedule. Since Diamer-Bhasha is expected to take 10 years to build (2011-2021), planners need to adjust their schedule estimate upwards to nearly 17 years (i.e. an actual opening date of 2028). Using our second approach, our multilevel Model 3 predicts that given that the dam's final decision to build was made in Pakistan by a democratically elected government, when the per capita income was USD 497 in 2000 constant dollars, a dam wall length of 998 meters, and an installed capacity of 4,500 MW, the expected outcome is a 60% schedule overrun. Thus, using either approach, Diamer-Bhasha can be expected to only open in 2027 when there remains a 20% risk of further delay. Pakistan is facing an energy crisis today (Kessides, 2011). A dam that brings electricity is 2027 will be a little late in coming.

Note, however, that if a dam of dimensions similar to Diamer-Bhasha were being built in the US (with its high per capita income of approximately USD 38,000), Model 3 predicts that it would face a schedule slippage of a mere 0.05%. Recall that per capita income is a useful proxy for the economic support that a country can provide for the construction of complex facilities. This suggests that rich and not developing countries best attempt very large energy projects, such as large dams. Even so, richer countries should adopt the risk management measures of the outside view illustrated here to choose prudently among energy alternatives.

Using their "inside" cost estimates, the net present benefits to cost ratio of the dam according to experts is 1.43 (WAPDA, 2011). Even assuming experts' calculations about potential benefits are accurate, although this is a doubtful assumption, the de-biased cost forecasts require an uplift of 44%-99% in constant prices suggest that the benefits to cost ratio will be below one. Diamer-Bhasha dam is a non-starter in Pakistan. This is without even discussing potential effects of inflation and interest rates, potential social and environmental costs, and opportunity cost Pakistan could earn by committing such vast amount of capital to more prudent investments.

Our reference class forecasting techniques suggests that other proposed large dam projects such as the Belo Monte, Myitsone, or the Gilgel Gibe III among many others in early planning stages are likely to face large cost and schedule overruns seriously undermining their economic viability. Large dams also exert an opportunity cost by consuming scarce resources that could be deployed to better uses, sinking vast amounts of land that could have yielded cash flows and jobs from agricultural, timber, or mineral resources. Risks related to dam safety, environment, and society further undermine viability of large dams. Decision-makers are advised to carefully stress test their proposed projects using the risk management techniques of the outside view proposed here before committing resources to them.

The outside view techniques applied to large dams have broader application in energy policy by helping public agencies (e.g. national planning and finance ministries, power and water authorities) private entrepreneurs and investors a framework to improve the upfront selection among alternatives. The problems of cost and schedule overrun are not unique to large hydropower dams. Preliminary research suggests that other large-scale power projects using nuclear, thermal, or wind production technologies face similar issues. Our research of large hydropower projects reveals that there is a serious dearth of valid and reliable data on the risk profiles of actually completed energy projects across the board. Much of the data in existing literature are drawn from surveys and interviews of dubious validity. At times, interest groups, seeking to promote a particular kind of scale or technology, also report distorted data. There is thus an urgent need to empirically document, in a comprehensive global database, the risk profiles of energy infrastructure assets of large, medium, and small scales across production technologies. For example, comparing the likely actual cost, schedule, and production volumes of a large hydropower dam project versus an on-site combined heat and power generator.

We propose that prior to making any energy investment, policy makers consult a valid and reliable "outside view" or "reference class forecast" (RCF) that can predict the outcome of a planned investment of a particular scale or production technology based on actual outcomes in a reference class of similar, previously completed, cases. Rigorously applying reference class forecasting to energy investments at various scales and production technologies will yield the following contributions:

- Create transparency on risk profiles of various energy alternatives,

not only from the perspective of financial cost and benefit but also environmental and social impact—hard evidence is a counter-point to experts' oft-biased inside view.

- Improve resource allocation through outside-in view to estimate costs, benefits, time, and broader impacts such as greenhouse gas emissions incurred in building a project and emission created or averted once a project becomes operational.

A comprehensive global dataset that can create such transparency on risk profiles of energy alternatives does not yet exist. We have sought to bridge this precise gap by providing impartial evidence on large hydropower dam projects. As a venue for further research we hope valid and reliable data on the actual cost, schedules, benefits, and impacts of other production technologies will become available to enable comparative analysis with novel implications for theory and practice.

**ENDNOTES**

[1] Cost overruns can also be expressed as the actual outturn costs minus estimated costs in percent of estimated costs.

[2] Note that the World Bank, Asian Development Bank, and the WCD typically report cost data in nominal U.S. dollars. We, however, converted these data, adapting methods from World Bank (1996: 85), into constant local currencies.

[3] Hufschmidt and Gerin (1970) report data on over 100 dams built in the United States between 1933-1967. The salient results of the study were that in nominal USD terms dams built by TVA suffered a 22 per cent cost overrun; U.S. Corps of Engineers overrun was 124 per cent for projects built or building prior to 1951, and 36 per cent for projects completed between 1951 and 1964; while U.S. Bureau of Reclamation overrun was 177 per cent for projects built or building prior to 1955 and 72 per cent for all projects built or building in 1960 (ibid: 277). Despite its large sample, Hufschmidt and Gerin (ibid) do not report data broken

down project-by-project. The validity and reliability of these data could not thus be established and were consequently excluded.

[4] A more comprehensive enquiry into planned versus actual benefits of dams is postponed until a future occasion but data available on 84 of the 186 large hydroelectric dam projects thus far suggests that they suffer a mean benefits shortfall of 11%.

[5] Recchia (2010, p. 2) explain further for why a R-squared measure cannot be used for a multilevel model because it unlike a single-level model which "includes an underlying assumption of residuals that are independent and identically distributed. Such an assumption could easily be inappropriate in the two[or multi]-level case since there is likely to be dependence among the individuals that belong to a given group. For instance, it would be difficult to imagine that the academic achievements of students in the same class were not somehow related to one another". Also see Kreft and Leeuw (1998) and Goldstein (2010).

[6] Note further that like in Model 1, the dependent variable in Model 3 is the inverse of schedule overrun (i.e. 1/x of the schedule overrun or Estimated/Actual schedule). Thus a negative sign on the Log of dam wall length (m) suggests that an increase in wall length decreases the inverse of the schedule overrun. In other words, increase in wall length increases schedule overrun.

**Appendix 1: Visual Representation of Model 1 (reported in Table 3)**

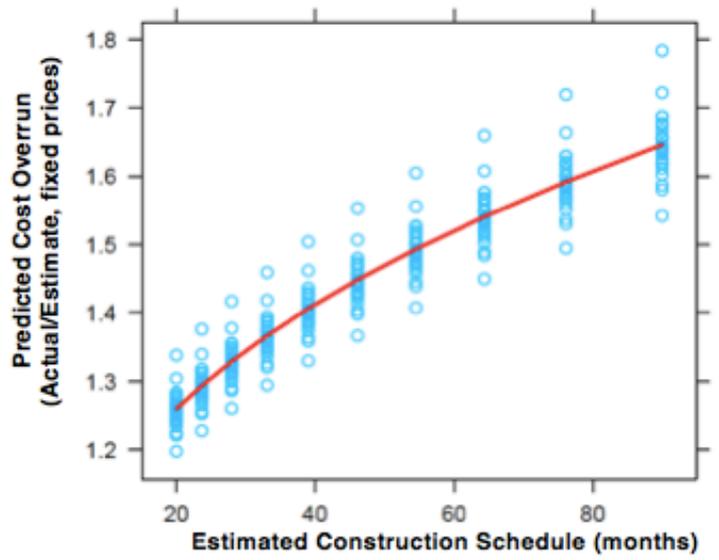

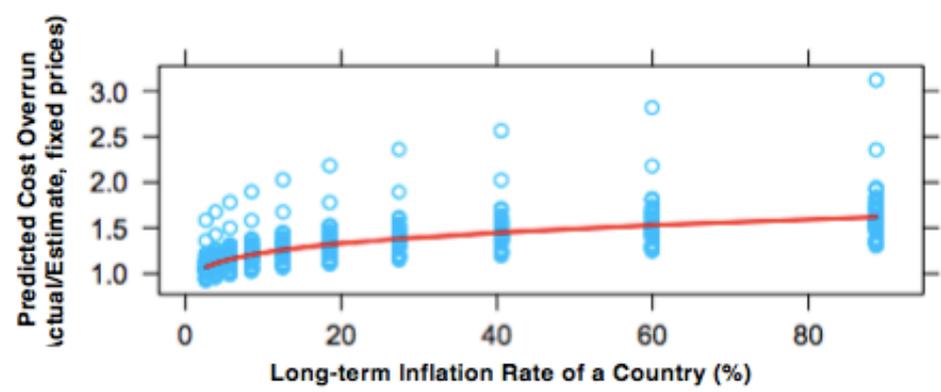